\pdfoutput=1

\documentclass[epj]{webofc}
\usepackage[utf8]{inputenc}
\usepackage[varg]{txfonts}   
\usepackage{booktabs}
\usepackage{xcolor}
\definecolor{darkred}{rgb}{0.4,0.0,0.0}
\definecolor{darkgreen}{rgb}{0.0,0.4,0.0}
\definecolor{darkblue}{rgb}{0.0,0.0,0.4}
\usepackage[bookmarks,linktocpage,colorlinks,
    linkcolor = darkred,
    urlcolor  = darkblue,
    citecolor = darkgreen]{hyperref}
\wocname{EPJ Web of Conferences}
\woctitle{Lattice2017}
\usepackage{amssymb}
\usepackage{amsmath}
\usepackage{graphicx}

\numberwithin{equation}{section}

\newlength{\dummysp}
\newcommand{\half}{\frac{1}{2}}

\newcommand{\beq}{\begin{eqnarray}}
\newcommand{\eeq}{\end{eqnarray}}
\newcommand{\nnn}{ \nonumber \\ }

\newcommand{\p}{{\partial}}

\newcommand{\s}{{\sigma}}

\newcommand{\vev}[1]{{\langle #1 \rangle}}

\newcommand{\ord}[1]{{{\cal O}(#1)}}
\newcommand{\gappeq}{\mathrel{\rlap {\raise.5ex\hbox{$>$}}
{\lower.5ex\hbox{$\sim$}}}}
\newcommand{\lappeq}{\mathrel{\rlap{\raise.5ex\hbox{$<$}}
{\lower.5ex\hbox{$\sim$}}}}
\newcommand{\myref}[1]{(\ref{#1})}

\newcommand{\ben}{\begin{enumerate}}
\newcommand{\een}{\end{enumerate}}
\newcommand{\bit}{\begin{itemize}}
\newcommand{\eit}{\end{itemize}}

\newcommand{\Ncal}{{\cal N}}

\newcommand{\Ocal}{{\cal O}}

\newcommand{\muhat}{{\hat \mu}}

\def\[{\left [}
\def\]{\right ]}
\def\({\left (}
\def\){\right )}

\begin{document}

\selectlanguage{english}

\title{Berezinskii-Kosterlitz-Thouless phase transition from lattice sine-Gordon model}

\author{
\firstname{Joel} \lastname{Giedt}\inst{1}\fnsep\thanks{Speaker, \email{giedtj@rpi.edu}. 
JG was supported in
part by the Department of Energy, Office of Science, Office of High Energy Physics,
Grant No.~DE-SC0013496.
Significant parts of this research were done using resources provided by the Open Science Grid 
\cite{osg1,osg2}, 
which is supported by the National Science Foundation award 1148698, and the U.S. Department of Energy's Office of Science.
We are also appreciative of XSEDE \cite{xsede} resources (Stampede),
where other significant computations for this study were performed.
}
\and
\firstname{James} \lastname{Flamino}
}

\institute{
Department of Physics, Applied Physics and Astronomy,
Rensselaer Polytechnic Institute, 110 8th Street, Troy NY 12180 USA
}

\abstract{
We obtain nonperturbative results on the sine-Gordon model using the lattice field technique.
In particular, we employ the Fourier accelerated hybrid Monte Carlo algorithm for our
studies.  We find the critical temperature of the theory based autocorrelation time,
as well as the finite size scaling
of the ``thickness'' observable used in an earlier lattice study by Hasenbusch et al.
}

\maketitle

\section{Introduction}
\begin{figure}
\begin{center}
\sidecaption
\includegraphics[width=3in]{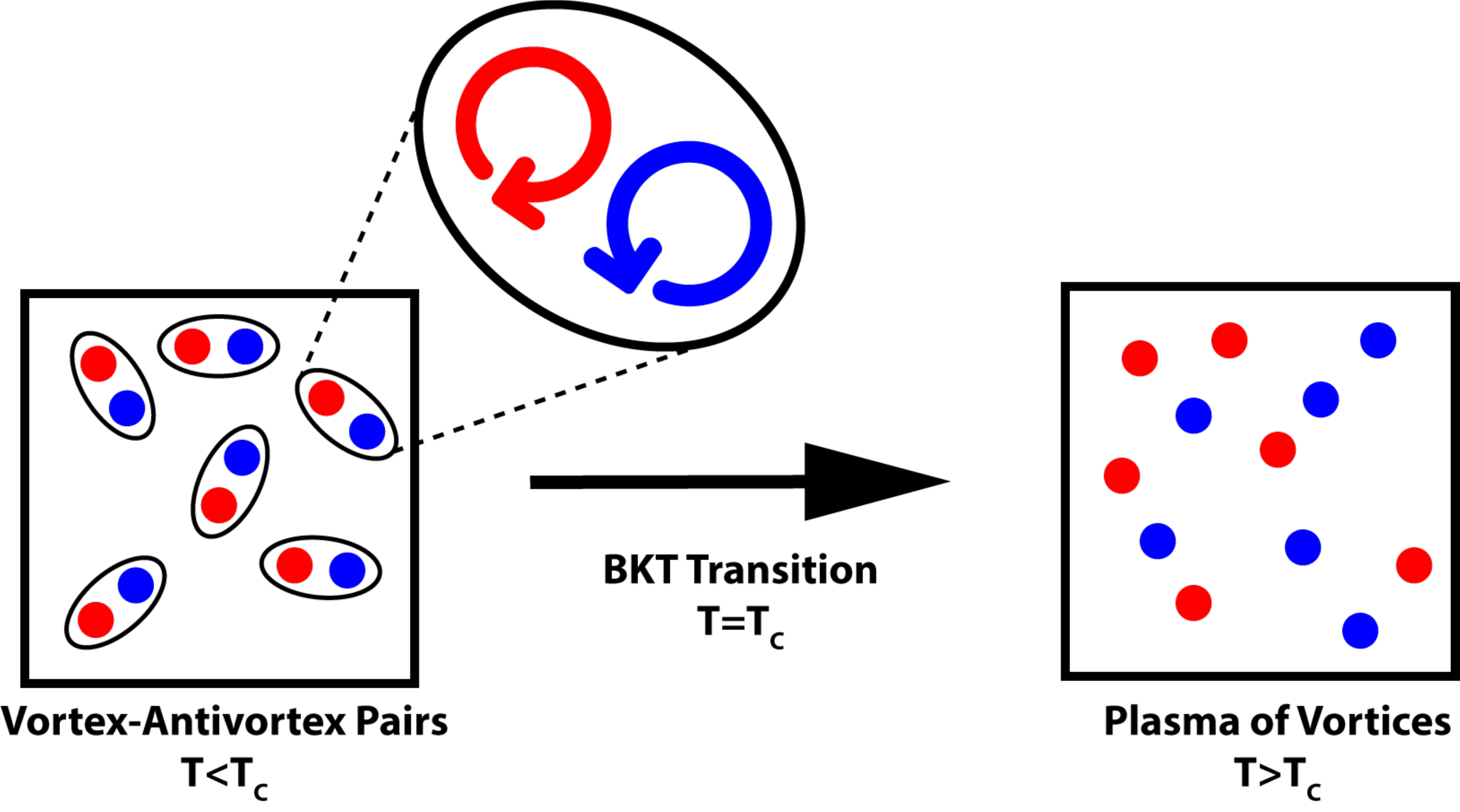}
\caption{Illustration of the vortex binding and unbinding on the two side of the BKT
transition. \label{bindingvorts} }
\end{center}
\end{figure}

Many revolutions in theoretical physics occured in the early 1970s; one of these was
the discovery that phase transitions did not always associate themselves with
spontaneous symmetry breaking and long-range order.  The physics of vortices,
and the corresponding topological phase transitions, allowed one to avoid the
theorems that opposed spontaneous symmetry breaking in two dimensions.  Besides
illustrating a defect driven topological phase transition, the related
systems, the XY model and the two-dimensional (2d) Coulomb gas, became
proving grounds for ideas related to the Wilsonian renormalization group.  We illustrate
the binding and unbinding of vortex -- anti-vortex pairs on
either side of the transition in Fig.~\ref{bindingvorts}.

The Berezinskii-Kosterlitz-Thouless (BKT) transition \cite{Berezinskii:1971,Kosterlitz:1973xp} was 
originally connected to the superfluid transition in two dimensions, such as the ${}^4$He thin
film.
Later it was realized that it could also occur in superconducting thin films, in spite
of being a charged superfluid.  In this case the supercurrents screen the fluctuations.
However, the screening length is given by $\Lambda = \lambda^2/d$, where $\lambda$ is
the magnetic penetration depth and $d$ is the film thickness.  Thus for large disorder
(hence large $\lambda$) and for sufficiently thin films, $\Lambda$ can still be large
enough for the algebraic order of the XY universality class to emerge.\footnote{See
for instance the review \cite{sgrvw1}.} 

The BKT transition is remarkably different from the usual Ginzburg-Landau
transition.  It is a transition without an order parameter.  There is a
line of conformal fixed points, rather than a single temperature
at which the theory is critical.  This entire low temperature region
displays algebraic ordering, i.e., correlation functions that are
power laws of the separation between operators.  Indeed, one can
regard it as a family of conformal field theories (CFTs), since the
anomalous dimensions (critical indices) are continuously varying with
temperatures.  In this sense the XY universality class can be regarded
as a two-dimensional (2d) toy model for $\Ncal=4$ super-Yang-Mills (SYM),
which also has continuously varying anomalous dimensions for (composite)
operators depending on the gauge coupling $g$ (or more generally, the
complexified coupling which incorporates the $\theta$ angle).

Another view of the sine-Gordon model is in terms of a description of vortices
that form in a 2d superfluid.  Here the circulation is quantized in units of
$h/M_A$, where $h$ is Planck's constant and $M_A$ is the atomic mass.  The
XY model angular variable $\theta(x)$ corresponds to this phase angle
of the superfluid wave function.

With the conventions that we use in this report,
the Euclidean action for the theory that we study is given by:
\beq
S[\phi] = \frac{1}{t} \int d^2 x ~ \left\{ \half [ \p_\mu \phi(x) ]^2 - g \cos \phi(x) \right\}
\label{sgact}
\eeq
Here, $t$ is the ``stiffness,'' which is identified with the inverse temperature in the corresponding
XY model, while $y \equiv g/2t$ is the fugacity of vortices.
The small $g$ behavior of this theory has been known for a long time.
For $t > 8\pi$ (the low temperature regime), the renormalization group (RG) flow (toward the infrared)
of $g$ is $g \to 0$.  We thus recover the theory
with long range correlations (algebraic ordering) in this part of the phase diagram.
Thus we see that $t > 8\pi$ corresponds
to the low temperature phase of the XY model, in
terms of its long distance behavior.  On the other hand, for $t < 8\pi$ (the high temperature regime), the
flow of $g$ is $g \to \infty$, which leads to screening and an
absence of criticality.

One should not confuse $t$ here with the reduced temperature $t_\text{red}=(T-T_\text{BKT})/T_\text{BKT}$.  Rather
$t$ maps to an inverse temperature in the XY model.  Nevertheless the BKT behavior $\xi \sim e^{a/\sqrt{t_\text{red}}}$
translates in the sine-Gordon theory into $\xi \sim e^{b/\sqrt{8\pi-t}}$ for $t < 8\pi$. 
Thus we retain the essential singularity at the critical temperature/stiffness.

\begin{figure}
\begin{center}
\sidecaption
\includegraphics[width=2.5in]{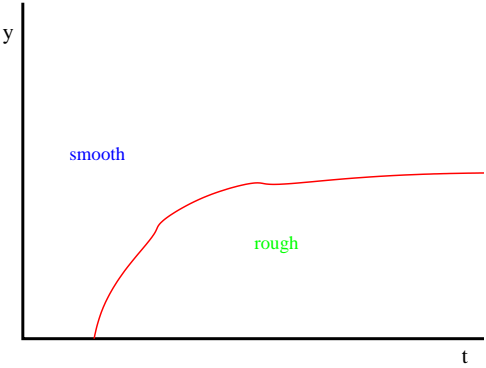}
\caption{Sketch of hypothetical phase boundary for sine-Gordon model.  \label{phasecurve} }
\end{center}
\end{figure}

This sine-Gordon model falls into a class of solid-on-solid (SOS) models, which display
the BKT transition. In the context of the SOS models, this is a roughening transition.
The relationship between roughening transitions in crystal facets and the XY universality
class was first described in \cite{ChuiWeeks76,ChuiWeeks78}.
The variable $\phi(x)$ is viewed as a height variable above a two-dimensional (2d)
surface---the facet of a crystal.  Hence above the critical $t$, where $\phi(x)$ becomes highly nonuniform,
it is described as the rough surface that grows at high temperatures.  
The critical line is described by $t_c(g)$, where $g$ then plays the role of
labeling different types of crystals (to the extent that a one parameter description is
usable).  However, when $y=g/2t \gg 1$ the potential term dominates and one is driven
to the uniform vacuum states where $\phi(x)$ is frozen to a multiple of $2\pi$.  Thus we
expect the curve $t_c(y)$ to tend to infinity as $y$ is increased, as shown in the
sketch, Fig.~\ref{phasecurve}.

\section{Fourier acceleration}
Here we describe the Fourier accelerated hybrid Monte Carlo (HMC) algorithm that is
used in our simulations \cite{Duane85,Ferreira93}.  The purpose of this modification of HMC is to reduce
autocorrelations between configurations of $\phi(x)$ that are produced in the
course of the simulation.  As a result, a shorter simulation can be run, while
producing comparable results and uncertainties.  Previous work using Fourier
accelerated HMC includes \cite{Espriu:1997jh,Catterall:2001jg}.

The HMC simulation proceeds as usual; the momentum field $\pi(x)$ is drawn
at random from a Gaussian distribution with unit variance.  This corresponds to
``integrating in'' a Gaussian field in the partition function, leading to the
``Hamilitonian''
\beq
H = \sum_x \bigg\{ \half \pi^2(x) - \frac{1}{2t} \phi(x) \Delta \phi(x) - \frac{g}{t} \cos \phi(x) \bigg\}
\eeq
where as usual the Laplacian is discretized by
\beq
\Delta \phi(x) = \sum_{\mu=1}^2 \p_\mu^* \p_\mu \phi(x) = \sum_{\mu=1}^2 \[ \phi(x+\muhat)
+ \phi(x-\muhat) - 2 \phi(x) \]
\eeq
In this notation $\p_\mu$ is the forward difference operator in the $\mu$ direction,
$\p_\mu^*$ is the backward difference operator, and $\muhat$ is a unit vector in the
$\mu$ direction.  We work in lattice units, $a=1$.  The Hamiltonian $H$
is evaluated to obtain $H(0)$.  Next, the fields $\pi(x),\phi(x)$ are
evolved according to Hamilton's equations
\beq
\dot \phi(x) = \frac{\p H}{\p \pi(x)}, \quad \dot \pi(x) = - \frac{\p H}{\p \phi(x)}
\eeq
for a trajectory of length $\tau$.  The numerical integration technique for this
evolution in fictitious time should be reversible and area-preserving, where area
refers to the functional integration measure $[d\pi(x) ~ d\phi(x)]$.  The standard
method is the leapfrog algorithm.  There is a step size $dt$ in this integrator,
and the Hamiltonian $H(\tau)$ at the end of the leapfrog trajectory will differ
from $H(0)$ due to error associated with not taking the $dt \to 0$ limit.  Thus
at the end of the trajectory, we apply the Metropolis accept/reject step
\beq
P_\text{acc} = \min(1,e^{-\Delta H}), \quad \Delta H = H(\tau) - H(0)
\eeq
to obtain a ``perfect'' algorithm, which will sample the functional integral with the
correct weight, the canonical distribution corresponding to $H[\pi(x),\phi(x)]$.

The Fourier acceleration enters into the leapfrog trajectory, where the
Fourier modes of $\phi(x)$ and $\pi(x)$ are integrated with a step size
\beq
dt(k) = dt/(\Delta(k) + m_\text{eff}^2)
\label{dtk}
\eeq
An equal number of steps $N_\tau = \tau/dt$ are taken for each mode $k$.
The Fourier transform of the force $-\p H/\p \phi(x)$ must also be used
in these equations.  In \myref{dtk}, $\Delta(k)$ is the Fourier transform
of $-\Delta(x,y)$:
\beq
\Delta(k) = \sum_{\mu=1}^2 4 \sin^2(\pi k_\mu / L_\mu), \quad k_\mu = 0,\ldots,L_\mu-1
\eeq
for an $L_1 \times L_2$ lattice.  By integrating the longer wavelength modes with a
larger step size, they are moved farther in configuration space.  This tends to reduce
autocorrelations, because it is precisely the long wavelength modes which are the source
of this difficulty.

We note that it is not necessary to take the $k$ dependent step size $dt(k)$ to be
of the form \myref{dtk}.  A follow-up study to the present one will be to optimize the
choice of $dt(k)$ using machine learning techniques.  The figure of merit that will
be maximized (i.e., the training goal) is the inverse of the integrated autocorrelation time.  The present
study is partly a preliminary step to this more extensive analysis of Fourier
acceleration, using the sine-Gordon model as a working context.  This is partly
motivated by the fact that two-dimensional real scalar field theories are easily
simulated on relatively small scale computing resources, and are thus well-suited
to exploratory studies.

\subsection{Thickness}
Following \cite{Hasenbusch:1994ef} we study the {\it thickness:}
\beq
\s^2=\frac{1}{V^2}\sum_{x,y} \vev{ (\phi(x)-\phi(y))^2 }
\eeq
Here $V = L \times L$ is the system size, ``volume.''  The thickness
 provides a measure of the roughness, on average, for a given parameter pair $(t,y)$.
This can be seen from the fact that if the entire lattice sits in
a single domain, with small fluctuations, then $\s$ will be small.
On the other hand, domain walls will contribute a nonzero result
even in the absence of fluctuations, so ground state disorder will
be picked up by the thickness observable.
In order to not bias toward a particular ground state, unless otherwise stated we begin all of our
simulations in studies of thickness with a
random start.

\begin{figure}
\begin{center}
\begin{tabular}{cc}
\includegraphics[width=2.5in]{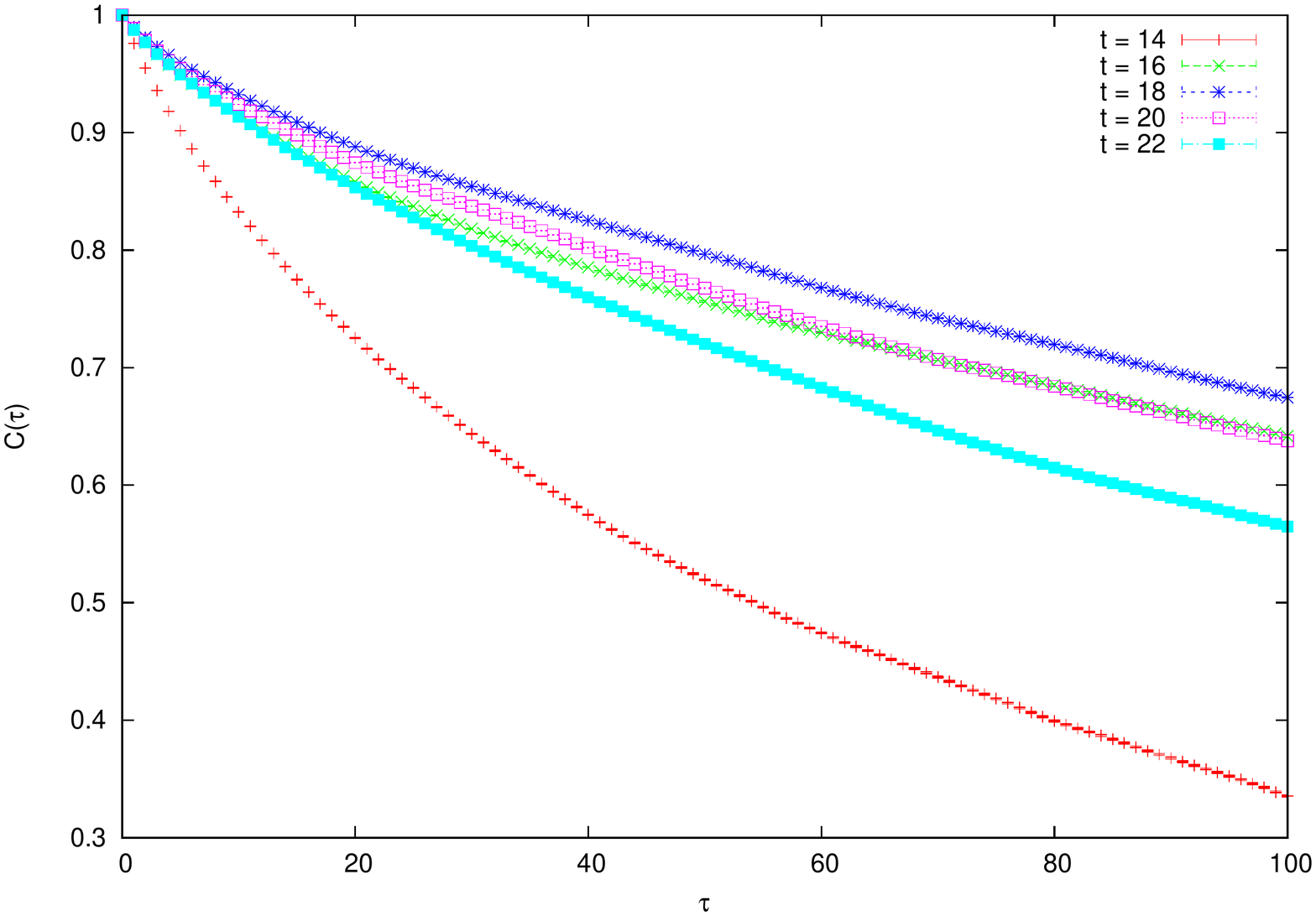} &
\includegraphics[width=2.5in]{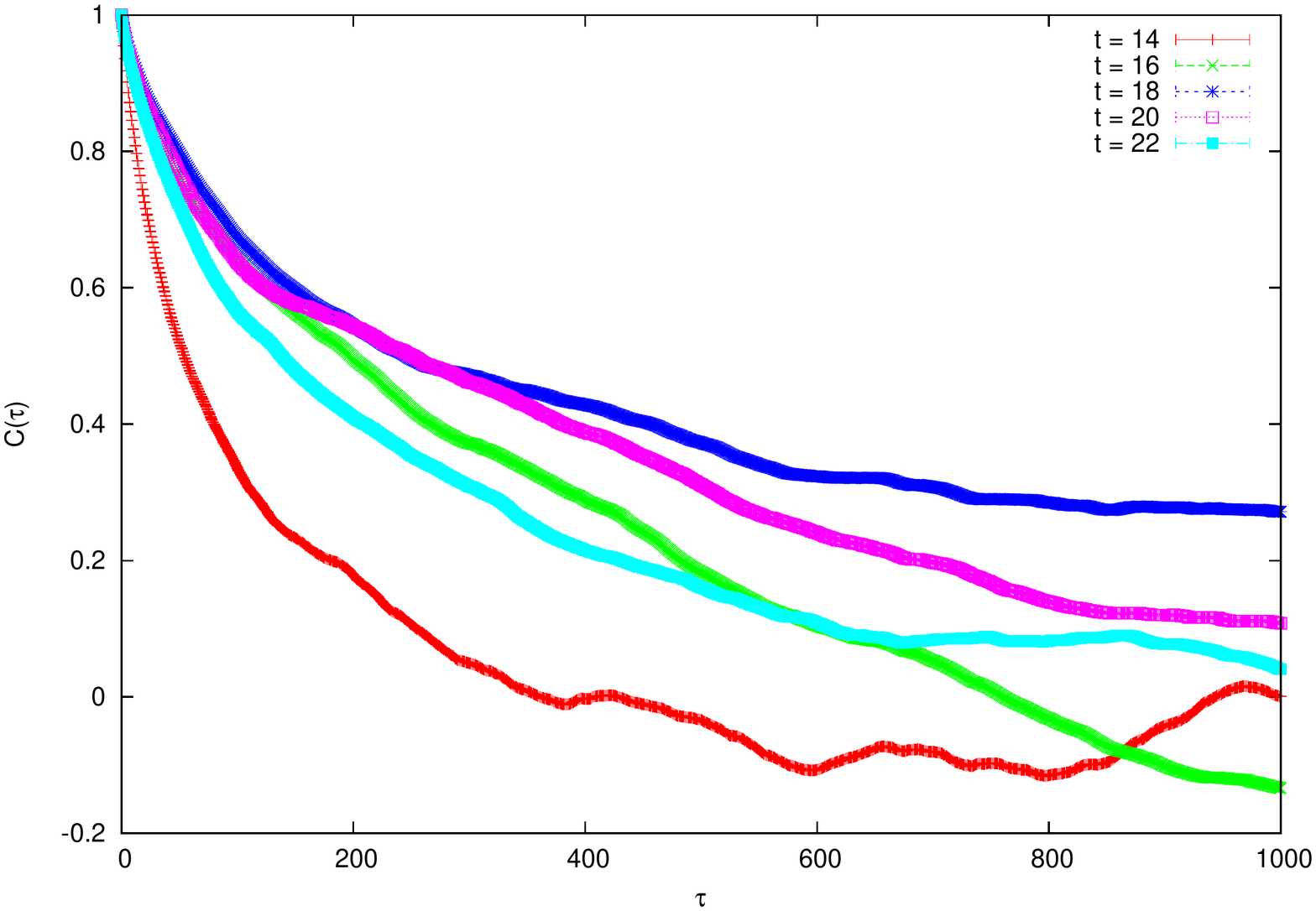} \\
\includegraphics[width=2.5in]{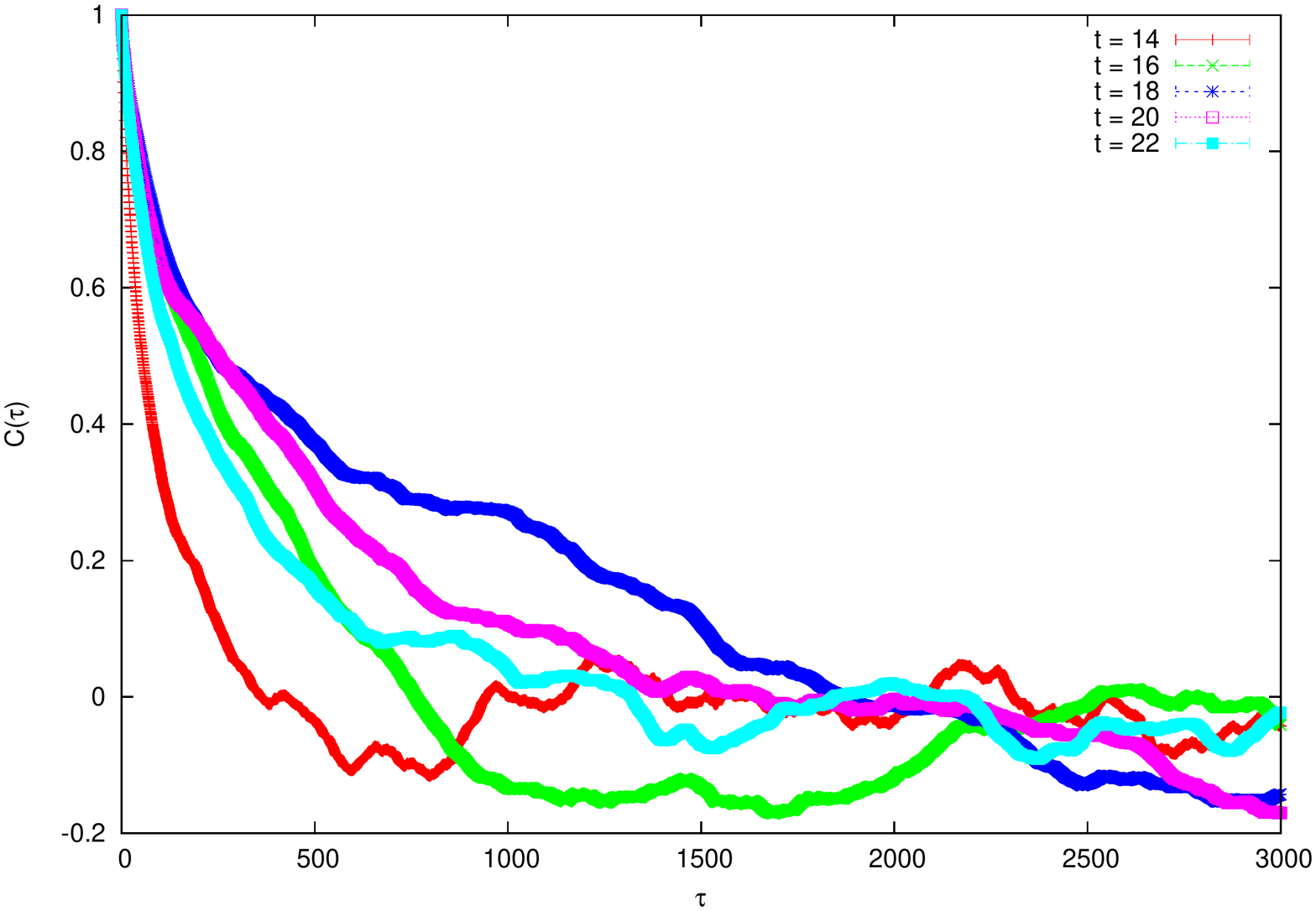} & 
\end{tabular}
\caption{The autocorrelation in the thickness observable
on a $64 \times 64$ lattice for $y=0.1$, and $t$ that
bracket the phase transition.  The three panels show progressively
longer simulation time scales.  It can be seen that for $t \approx t_c \approx 18$,
the autocorrelation is the greatest, as is to be
expected.  It can be seen that even after 1,000 updates, autocorrelations
are still significant for temperatures close to the
critical temperature of $t_c \approx 18$, whereas for $\ord{3000}$ time
steps all memory of the initial configuration has vanished.
\label{acthick_all}}
\end{center}
\end{figure}

It is important to understand the autocorrelation for 
this quantity before taking any averages and estimating uncertainties.
This is because we need several autocorrelation times in
order to thermalize (i.e., to sufficiently advance the memory
kernel beyond the initial conditions), and we need to know the
separation between statistically independent samples---where the
Monte Carlo simulation is effectively a Markov process.
The autocorrelation for any observable $\Ocal(t)$, where
$t$ here is the simulation time (measured in molecular dynamics time units), is
given by
\beq
C(t) &=& \frac{1}{{\cal N}} \frac{1}{N-t} \sum_{t' = 0}^{N-t-1} ( \Ocal(t'+t)-\vev{\Ocal} ) ( \Ocal(t')-\vev{\Ocal} ) \nnn
{\cal N} &=& \frac{1}{N} \sum_{t=0}^{N-1} (\Ocal(t) - \vev{\Ocal})^2,
\quad \vev{\Ocal} = \frac{1}{N} \sum_{t=0}^{N-1} \Ocal(t)
\eeq
and $N$ is the total number of time steps in the simulation.  In the
present case,
\beq
\Ocal = \frac{1}{V^2} \sum_{x,y} ( \phi(x) - \phi(y) )^2
\eeq
In Fig.~\ref{acthick_all}
we show short, long and very long time scales.  It can be
seen that there is an initial rapid decay, but that
a longer time component also contributes.  In
fact it takes $\ord{10^3}$ or more updates to obtain a completely
independent configuration.  These results are for $y=0.1$
with $t$ values that bracket what will turn out to be the
critical temperature, $t_c \approx 18$.  In fact, it can
be seen that $t \approx t_c$ yields the longest
autocorrelation times, which is to be expected.  This is
because critical fluctuations, which have very long
wavelength, lead to significant slow-down in typical
Monte Carlo algorithms.  We see that the Fourier acceleration
has not been entirely effective in alleviating this
critical slowing down.  On the other hand, it can be
seen that monitoring the autocorrelation can be a surprisingly
good way to locate the critical temperature.

\begin{figure}
\begin{center}
\begin{tabular}{cc}
\includegraphics[width=2.5in]{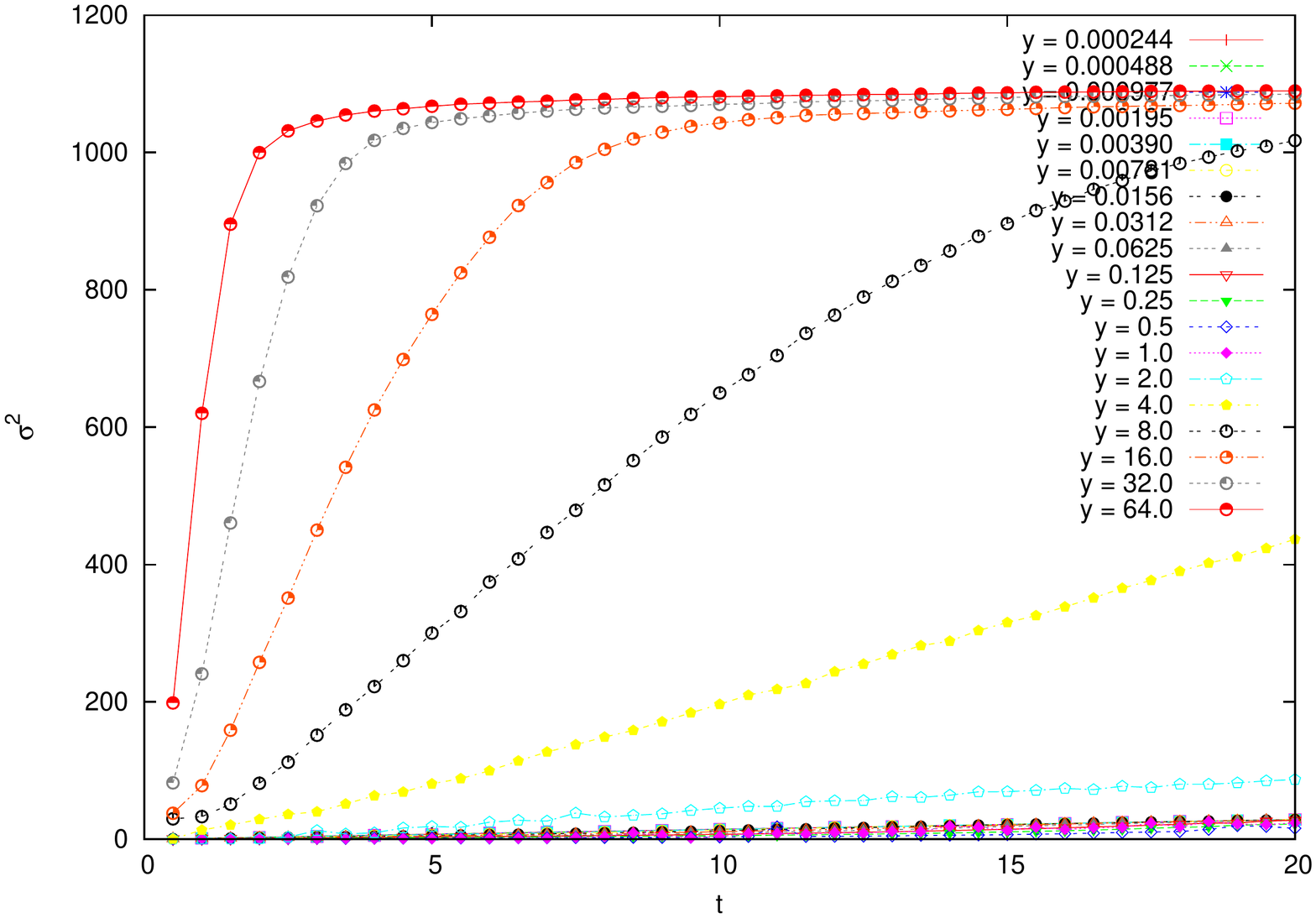} &
\includegraphics[width=2.5in]{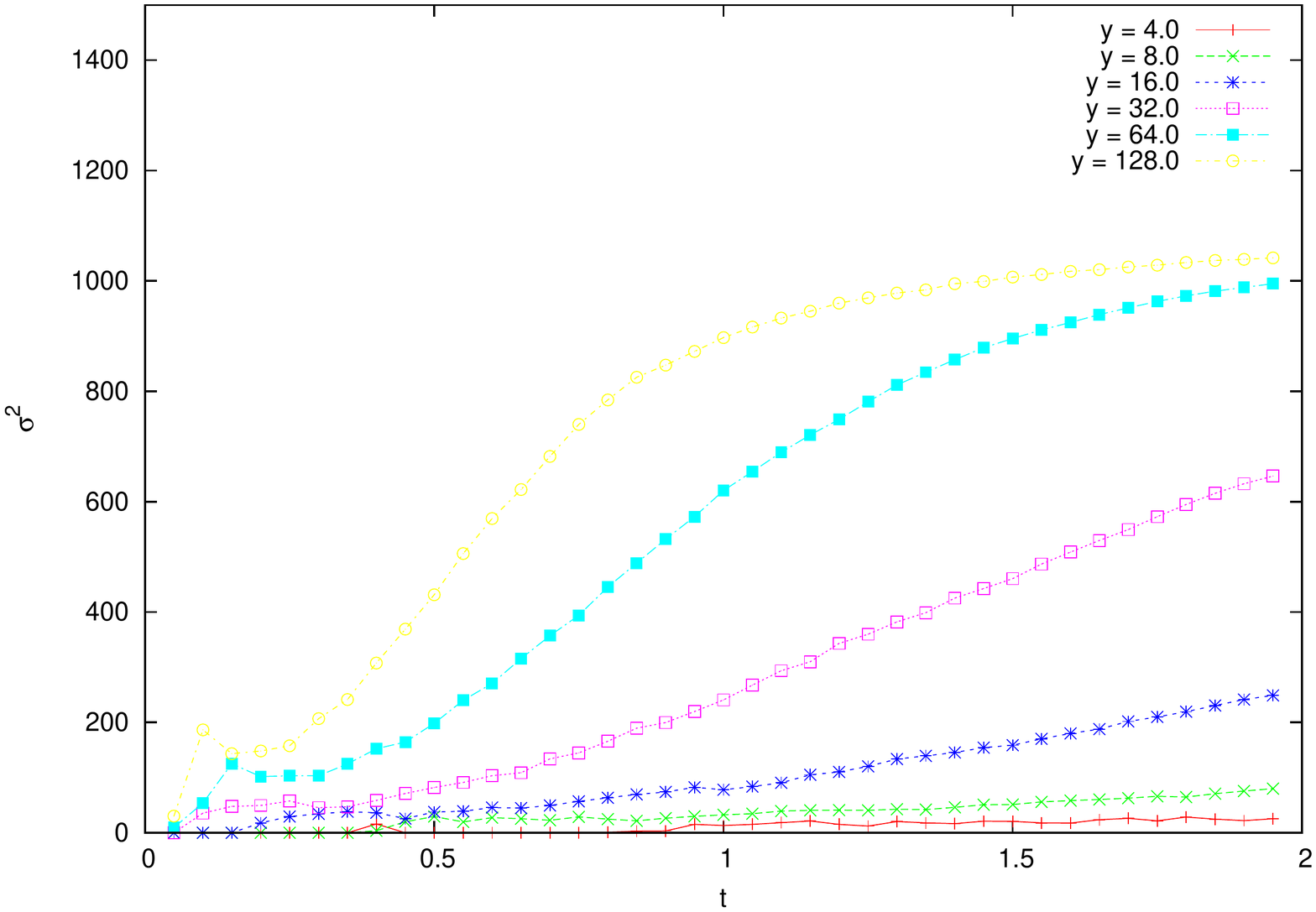}
\end{tabular}
\caption{Thickness versus $t$ for various $y$ for an $L=64$ lattice.
It can be seen that a cross-over between different behaviors occurs for $y \sim 1$,
with a nonlinear dependence of $\s^2$ on $t$ appearing for large $y$.
On the right, we show thickness versus $t$.
Some interesting features emerge at the smallest values of $t$ seen in this resolution,
though at the smallest value of $t$ in the right-hand panel, numerical
instabilities may be setting in.
\label{thvtyscan_both}}
\end{center}
\end{figure}

It is interesting to hold $y = g/2t$, the fugacity, fixed as we vary $t$.    
In Figs.~\ref{thvtyscan_both} it can be seen that
for small values of $y$ and $t$,
the thickness just behaves as a Gaussian variance directly proportional
to $t$.  It thus appears that the $y$ coupling has essentially no effect in this
regime, other than to determine the slope of the line.  
At larger values we begin to see nonlinear effects.

\begin{figure}
\begin{center}
\sidecaption
\includegraphics[width=3in]{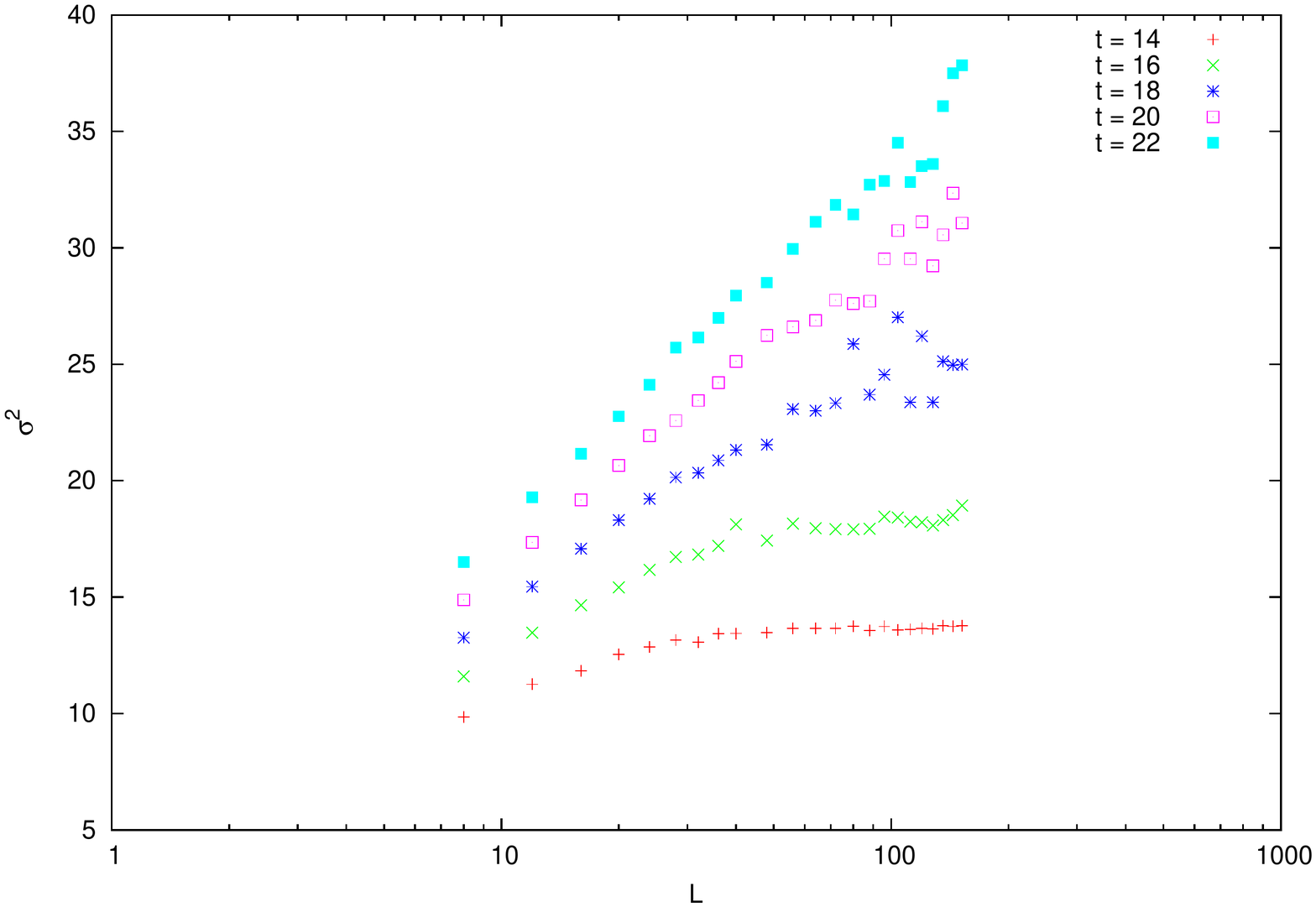}
\caption{The thickness observable transitioning between the
two behaviors \myref{bigLpred} at $t_c \approx 18$.  
Here, we have used $y=0.1$. 
\label{transition} }
\end{center}
\end{figure}

Hasenbusch et al.~provide a perturbative prediction for the finite size scaling
of the thickness above and below the transition temperature \cite{Hasenbusch:1994ef}.
In our conventions it is given by
\beq
\s^2 \simeq \left\{ 
\begin{array}{cc} 
\frac{t}{\pi} \ln L + \text{const.}, & \quad t > t_c \\
\text{const.} & \quad t < t_c
\end{array}
\right.
\label{bigLpred}
\eeq
in the large $L$ limit.  They have also verified this behavior in Monte Carlo
simulations using a cluster algorithm.  We conduct the same study, but with the
Fourier accelerated HMC.  Fig.~\ref{transition} shows precisely this behavior, with $t_c \approx 18$.  We
note that the $y \to 0$ limit (fugacity expansion) predicts $t_c = 8\pi \approx 25$, so there is
apparently some renormalization of $t$ arising from ${\hat y} = 0.1$.  If we fit the coefficient $c$ of 
$\s^2 \simeq c \ln L + \text{const.}$
for the $t=22$ data, we obtain $c \approx 6.7 \pm 0.2$, which is to be compared with
the results of \cite{Hasenbusch:1994ef} which predict $t/\pi = 7.0$.  
We attribute the small discrepancy ($1.5\s$) to a possible underestimation of errors and
a renormalization of $t$.  Once differences in conventions are taken into account,
our result of $t_c \approx 18$ is also in agreement with \cite{Hasenbusch:1994ef}.

\section{Outlook}
In a forthcoming full length article \cite{forthcoming} we will present further
results of our simulations, including clustering of the field $\phi(x)$
into domains and the entropy of the system.

Along the line of fixed points that occurs at $t > 8\pi$, we will have a conformal
field theory describing the infrared physics.  Since this is a two-dimensional theory,
the conformal group is infinite dimensional, represented in terms of the
Virasoro algebra.  One of the things that we would like to eventually investigate is the
emergence of this infinite dimensional symmetry algebra at long distances.
Indeed, it is not a simple map to find the field operators of the ultraviolet
theory that will correspond to the Virasoro generators.  This is because the
symmetry is not a property of the ultraviolet theory, due to the nonzero $g$.

Various methods can be employed to improve the simulation.  This includes
embedding a cluster algorithm such as was done in \cite{Hasenbusch:1994ef},
as well as parallel tempering to allow better sampling of the ground
state degeneracy.  These advances will appear in our future works.

\clearpage

\bibliography{latproc}

\end{document}